\documentclass[article,onecolumn,amsmath,amssymb,aps,
]{revtex4}

\usepackage{epsfig,bm,epsf,graphics}
\usepackage{graphicx}
\usepackage{dcolumn}
\usepackage{bm}
\usepackage{xcolor}

\begin{document}
\preprint{APS/123-QED}

\title{PHYSICS OF 2D MAGNETS AND MAGNETIC THIN FILMS:\\
SURFACE STRUCTURE AND PHASE TRANSITIONS,  CRITICALITY AND SKYRMIONS}

\author{ Hung T. Diep\footnote{diep@cyu.fr, corresponding author}}
\affiliation{%
 Laboratoire de Physique Th\'eorique et Mod\'elisation,
CY Cergy Paris Universit\'e, CNRS, UMR 8089\\
2, Avenue Adolphe Chauvin, 95302 Cergy-Pontoise Cedex, France.\\
}



\date{\today}


\begin{abstract}
Recently, there is an increasing renewed interest in 2D magnetism such as  Van der Waals magnets. The physics of 2D magnetism and ultra-thin magnetic films has a long history. This chapter is a review devoted to some fundamental theoretical properties of 2D magnets and and magnetic thin films including frustrated systems and topological  spin textures. These properties allow to understand macroscopic behaviors experimentally observed in thin films and superlattices where the surface and the interface play a crucial role. The chapter begins with a review on 2D magnets, their spin structures and phase transitions. Next, the case of thin films is considered. The theory of surface spin waves is discussed  in various situations with and without surface reconstruction of spin ordering. Various interactions are taken into account: surface interaction different from the bulk one, competing interactions, Dzyaloshinskii-Moriya interaction.  Surface phase transitions are shown in some particularly striking cases.  Finally, some cases of topological spin textures called "skyrmions" are reviewed. All the results shown in this chapter have been published in various research papers cited in the text.  Therefore, we will discuss some important results but avoid to enter complicated methods. Instead, the reader is referred to original papers for detailed demonstrations. 
\vspace{0.5cm}
\begin{description}
\item[PACS numbers: 5.10.Ln;64.30.+t;75.50.Cc]
\end{description}
\end{abstract}


\maketitle


\section{Introduction}

Investigations on properties of nanoscale magnetic systems have been intensively carried out in the last five decades. This is because there was a search for smaller and smaller systems with high capacities of memory stockage. Magnetic thin films and nanodots are among the objects among the most studied experimentally, numerically and theoretically.\cite{zangwill,bland-heinrich,DiepTM,Binder-surf,Diehl}  In the 80's, the discovery of giant magneto-resistance (GMR) has led to a generation of high-density hard disks.\cite{Fert,Grunberg}  Intensive researches continue to look for new effects, new phenomena which can lead to even more efficient, energy-saving electronic applications. Spintronics is one of the future technologies which is based on spins or groups of spin.  A single spin with up and down orientations represents the smallest entity for a computer binary bit or logic gates in principle, but how to control the orientation of such a spin? Recent research focuses on groups of a few dozens of spins having the double-fold states such as nanodots \cite{ABR2021} and skyrmions \cite{Zhang} which have left or right chirality. With a magnetic field, for example, one can change its chirality. In multiferroic materials, the chiralily of skyrmions can be controlled by an electric field \cite{Sharafullin2019}. 

Since the discovery of the effect of the frustration in magnetic materials, an enormous number of investigations has been realized, theoretically, experimentally and numerically. The reader is referred to Refs. \cite{Binder-surf,Diehl,DiepFSS} for reviews on various topics.

In this chapter, we show some examples of  magnetic mono-layers or films where striking surface effects have been discovered. We  show surface states, surface phase transitions and other properties that may have applications in spintronics. The criticality of thin films is presented. We show that this is an important subject which may allow us to understand critical exponents experimentally observed in thin films. We also review  important works since 2003 \cite{Bogdanov2003} on what is called "skyrmion" (topologically protected spin structure containing a dozen of spins).  We recall theoretical and experimental works on "skyrmion crystal" which is  a "superstructure" of skyrmions periodically arranged in a lattice structure. 

Since the aim of this chapter is to review some striking aspects of the surface magnetism, we shall show main results and discuss on their physical implications. We shall not show complicated technical details in order to make  the reading easy. We will refer the reader to original papers for the mathematical and computing techniques.  

The chapter is organized as follows. In section \ref{Frustration} we briefly present the notion of frustation which allows to understand the following sections. In section \ref{SW} we consider the case of thin films where striking surface properties are reviewed. We discuss here both non  frustrated and frustrated films. In section \ref{critical} we show results on the criticality of thin films. In section \ref{DMSW} we calculate the spin-wave spectrum in a monolayer with Dzyaloshinskii-Moriya interaction.  In section \ref{Skyrmions} we describe skyrmions resulted from the Dzyaloshinskii-Moriya interaction. Concluding remarks are given in section \ref{Concl}


\section{Frustration}\label{Frustration}

A system is said "frustrated" when a number, or all,  of interactions between two spins are not fully satisfied.  This situation is encountered when there are a competition between various interactions in the system. It can also happen when there is only one kind of interaction in  the system, but the geometry of the lattice does not allow for a full satisfaction of that interaction everywhere. This is the case of a triangular lattice with an antiferromagnetic interaction between nearest neighbors (NN), or the face-centered cubic (FCC) lattice and the hexagonal-closed-packed (HCP) lattice, with antiferromagnetic NN interaction. We call these case  "geometry frustration".  

We focus in this paper on the case of frustrated systems in two dimensions (2D) and in thin films.  For the highly frustrated Ising models, there has been a large number of exactly solved models in 2D. We can mention a few of them in 
Refs. \cite{KaNa,Vaks,Aza87,Diep91b,Diep91a,Diep92,Diep92a}. Other models of decorated frustrated Ising lattices have also be exactly solved \cite{Strec1,Strec2,Strec3}. The reader is referred to chapter 1 of Ref. \cite{DiepFSS} for a detailed description of these exact methods.  Let us enumerate a few striking effects due to the frustration in these axactly solved systems:

(i) the ground state (GS) degeneracy is very high, in many cases the GS degeneracy is infinite,

(ii) there exists a reentrance phase (paramagnetic phase) between two ordered phases when the temperature $T$ increases.  The low-$T$ ordered phase is ferromagnetic F (or antiferromagnetic, depending on the model). The high-$T$ ordered phase X is in fact a partially ordered phase: one part (a sublattice) is disordered while the remaining part is ordered. One assists here a coexistence between order and disordered (see for example Fig. 3 of Ref. \cite{Aza87}),

iii) in the reentrance phase, there exists a disorder line, with or without dimension reduction, which separates the pre-ordering fluctuations between phases F and X (see Refs. \cite{Ste1,Ste2,Ste3} for the meaning of the disorder lines),

iv) some models exhibit successive phase transitions (we in some models found up to 5 transitions while increasing $T$) and unusual forms of disorder lines (see chapter 1 of Ref. \cite{DiepFSS})

In the case of vector spins (XY and Hesenberg spins), the GS spin configuration is non-collinear, unlike the collinear configuration in ferromagnets and antiferromagnets.  The well-known example is the 120-degree spin structure in the antiferromagnetic triangular lattice.  Note that the helimagnetic structure has been determined a long time ago \cite{Yoshimori,Villain59}.  The nature of the phase  transition in the vector spin cases is often difficult to determine. To our knowledge, the phase transition in most (if not all) frustrated systems in 3D is of the first-order transition. Well-known examples are the cases of HCP antiferromagnets \cite{Diep1992HCP,Hoang-Diep2012HCP} and the FCC antiferromagnets \cite{Diep-Kawamura}.
The antiferromagnetic stacked triangular lattice is perhaps the most controversial during more than 30 years as seen in Refs. \cite{Kawamura1988,Kawamura1998,Delamotte,Delamotte2010,Ngo-Diep,Ngo-DiepXY,Reehorst,Delamotte2024}. The most recent works among these papers are in favor of the first-order character of the transition of this system.

Let us examine some simple cases in the following.

\subsection{Definition}\index{frustration, definition}

Uniformly frustrated systems are periodicallly defined: the translational invariance allows in many cases to consider only an elementary cell of the lattice in
search for the system GS. 
The lattice cell is in general a polygon
formed by faces hereafter called "plaquettes". For example, the
elementary cell of the simple cubic (SC) lattice is a cube with six
square plaquettes, the elementary cell of the FCC lattice is a
tetrahedron formed by four triangular plaquettes. Let $J_{i,j}$ be
the interaction between two NN spins of the plaquette. According
to the definition of Toulouse,\cite{Tou} the plaquette is
frustrated if the parameter $P$ defined below is negative
\begin{equation}
P=\prod_{\left<i,j\right>}\mathrm{sign}(J_{i,j}), \label{frust1}
\end{equation}
where the product is performed over all $J_{i,j}$ around the
plaquette. Two examples of frustrated plaquettes are i) a triangle with three antiferromagnetic bonds
and a square with three ferromagnetic bonds and one
antiferromagnetic bond.  $P$ is negative in both cases.  If one tries to put Ising spins on those plaquettes, 
one of the bonds around the  plaquette will not be satisfied. For
vector spins, we show below that in the lowest energy state, each
bond is only partially satisfied.

For the triangular plaquette, the degeneracy is
three, and for the square plaquette it is four in the Ising case, in addition to the
degeneracy associated with global spin reversal.  Therefore, the
degeneracy of an infinite lattice composed of such plaquettes is
infinite, unlike the ferro- and antiferromagnetic cases.

Hereafter, we analyze the GS of a few frustrated systems with XY and Heisenberg
spins.

\subsection{Non collinear spin configurations}\index{non-collinear spin configuration}

In the case of $XY$ spins, one can calculate the GS configuration of the triangular antiferromagnet
by minimizing the energy of the plaquette $E$ while keeping the
spin modulus constant. In the case of the triangular plaquette,
suppose that spin $\mathbf S_i$ $(i=1,2,3)$ of amplitude $S$ makes
an angle $\theta_i$ with the $\mathbf {Ox}$ axis. Writing $E$ and
minimizing it with respect to the angles $\theta_i$, one has
\begin{eqnarray}
E&=&J(\mathbf S_1\cdot \mathbf S_2+\mathbf S_2\cdot \mathbf S_3+\mathbf S_3\cdot \mathbf S_1)\nonumber \\
&=&JS^2\left[\cos (\theta_1-\theta_2)+\cos (\theta_2-\theta_3)+
\cos (\theta_3-\theta_1)\right],\nonumber \\
\frac{\partial E}{\partial \theta_1}&=&-JS^2\left[\sin
(\theta_1-\theta_2)-
\sin (\theta_3-\theta_1)\right]=0, \nonumber \\
\frac{\partial E}{\partial \theta_2}&=&-JS^2\left[\sin
(\theta_2-\theta_3)-
\sin (\theta_1-\theta_2)\right]=0, \nonumber \\
\frac{\partial E}{\partial \theta_3}&=&-JS^2\left[\sin
(\theta_3-\theta_1)- \sin (\theta_2-\theta_3)\right]=0. \nonumber
\end{eqnarray}

A solution of the last three equations is
$\theta_1-\theta_2=\theta_2 -\theta_3=\theta_3-\theta_1 =2\pi/3$.
This solution can be also obtained by writing the following equality
$$
E=J(\mathbf S_1\cdot \mathbf S_2+\mathbf S_2\cdot \mathbf
S_3+\mathbf S_3\cdot \mathbf S_1)
=-\frac{3}{2}JS^2+\frac{J}{2}(\mathbf S_1 + \mathbf S_2 + \mathbf
S_3)^2.
$$
The minimum of $E$ corresponds to $\mathbf S_1 + \mathbf
S_2 + \mathbf S_3=0$ which yields, using a geometric construction, the $120^\circ$
structure. This is true also for Heisenberg spins.

We can do the same calculation for the case of the frustrated
square plaquette. \cite{Berge} Suppose that the antiferromagnetic bond connects
the spins $\mathbf{S}_1$ and $\mathbf{S}_2$. We find
\begin{equation}
\theta_2-\theta_1=\theta_3 -\theta_2=\theta_4-\theta_3
=\frac{\pi}{4} \textrm { and }\theta_1-\theta_4=\frac{3\pi}{4}
\label{frust2}
\end{equation}

In the general case where the antiferromagnetic bond is equal to $-\eta J$, the solution
for the angles is\cite{Berge}
\begin{equation}
\cos\theta_{32}=\cos\theta_{43}=\cos\theta_{14}\equiv \theta=\frac
{1}{2}[\frac {\eta+1}{\eta}]^{1/2}\label{frust2a}
\end{equation}
and $|\theta_{21}|=3|\theta|$, where $\cos\theta_{ij}\equiv
\cos\theta_{i}-\cos\theta_{j}$.

The above solution exists if $| \cos\theta |\leq 1$, namely
$\eta>\eta_c=1/3$.  

\begin{figure}[ht]
\centering
\includegraphics[width=3.2 in]{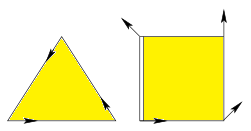}
\caption{ Non-collinear spin configuration of frustrated triangular and square
plaquettes with $XY$ spins.  In the triangular case, all bonds are antiferromagnetic. 
In the square lattice case the ferro- and antiferromagnetic
interactions  $J$ and $-J$ are shown by single and double lines, respectively.}
 \label{fig:IntroNCFT}
\end{figure}

We show the frustrated triangular and square lattices in Fig.
\ref{fig:IntroNCFT} with $XY$ spins.

There is  a two-fold degeneracy  resulting from
the mirror symmetry  with respect to an axis, for
example the $y$ axis in Fig. \ref{fig:IntroNCFT}. Therefore the
symmetry of these plaquettes is of Ising type O(1), in addition to
the symmetry SO(2) due to the invariance by global rotation of the
spins in the  plane. The nature of the phase transition in the lattice generated by  the above plaquette 
has been found  to be a coupled Ising-XY universality class \cite{BoubcheurDiep1998,Nightingale,Granato,Lee,Granato1}.

Another example is the case of a chain of Heisenberg spins with
ferromagnetic interaction $J_1(>0)$ between NN and
antiferromagnetic interaction
 $J_2 (<0)$ between NNN. When
$\alpha = |J_2|/J_1$ is larger than a critical value
$\alpha_c$, the GS spin configuration is non-collinear. The turn angle is given by

\begin{equation}
\cos \theta=\frac{J_1}{4|J_2|} \longrightarrow \theta= \pm \arccos
\left(\frac{J_1}{4|J_2|}\right).
\end{equation}
This solution is possible if $-1\le \cos\theta \le 1$, i.e.
$J_1/\left(4|J_2|\right)\le 1$ or $|J_2|/J_1\ge 1/4 \equiv
\alpha_c$.

\begin{figure}[ht]
\centering
\includegraphics[width=3.2 in]{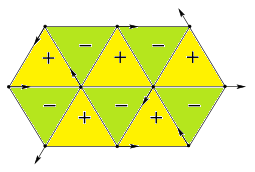}
\caption{\label{fig:IntroAFTL} Antiferromagnetic triangular
lattice with  $XY$ spins. The positive and negative chiralities
are indicated by $+$ and $-$.}
\end{figure}

By translating the triangular  plaquettes  shown in Fig.
\ref{fig:IntroAFTL}, we obtain the GS spin configuration of the antiferromagnetic triangular lattice. One sees 
that the GS corresponds to the state where all
triangle plaquettes of the same orientation have the same chirality:
plaquettes $\bigtriangleup$ have positive chirality
and plaquettes $\bigtriangledown$ have negative chirality.  Mapping the chiralities into the Ising spins, we have a perfect
antiferromagnetic Ising order. This order is broken at a phase
transition temperature where the chirality ordering (or Ising staggered magnetization) vanishes.

\section{Surface states: surface spin waves, surface spin structures, surface phase transition}\label{SW}
Let us take for example a thin film. The spins on the surfaces have less neighbors than those in the bulk. This causes many effects: the electronic orbitals, the exchange interactions, the spin configuration, the lattice constants,... are more or less modified.  As consequences, the electronic structures and the spin-wave spectrum may have what we call "surface states". These states lead to macroscopic properties of the magnet different from the bulk ones such as the surface magnetization and the surface phase transition.  
In the following, we will show the link between the existence of a surface and those striking surface properties.

\subsection{Surface spin waves}

When a crystal is infinite in all directions, spin waves are excited due to the interactions between spins on the lattice. A spin-wave is characterized by a wave vector $\mathbf k$ which have three components in three lattice directions. The spin-wave energy $E=\hbar \omega$ depends on  $\mathbf k$, $\omega$ being the spin-wave frequency.  The spin-wave spectrum is given by $E=\hbar \omega (\mathbf k)$ in the $k$-space defined  inside the first Brillouin zone. Now, if the crystal is semi-infinite or a thin film, the values of  $\mathbf k$ in the direction perpendicular to the surface are no more determined by the periodic condition of the lattice, namely the Fourier transform cannot be done in that direction. As a consequence, surface spin waves can be observed: these waves propagate parallelelly to the surface but are damped when propagating  toward inside the crystal (or film). They are called "surface-localized spin waves". We have studied in the past the existence of surface spin waves in a BCC ferromagnetic and antiferromagnetic thin films \cite{DiepLevyNagai,diep81} using the Green's function method for collinear spin configurations. We have also studied surface spin waves in  ferrimagnetic semi-infinite crystal \cite{Diep75}. The reader is referred to those references for details.  

\subsection{Surface non-collinear spin structures}
 In the case where there are competing interactions, namely ferromagnetic and antiferromagnetic interactions simultaneously acting on a spin, there are often a surface reconstruction which shows a surface spin ordering different from the bulk spin ordering. We give some examples in the following. 

{\bf Example 1:} Helimagnetic spin structure in a thin film:

This system has been studied in Ref. \cite{DiepHeli}.  We have determined the surface spin structure and used the Green's function method for non-collinear spin configuration \cite{QuartuDiep} to calculate spin-wave spectrum. The magnetization of each film layer has been calculated as a function of temperatute ($T$), using the spin-wave spectrum. This is briefly shown below.
Low-temperature properties in helimagnets such as spin-waves \cite{Harada,Rastelli,Diep89,Quartu1998} and
heat capacity \cite{Stishov} have been extensively investigated. Helimagnets belong to a class of frustrated vector-spin systems.  Note some works have been devoted to helimagnets: surface spin structures \cite{Mello2003}, MC simulations \cite{Cinti2008}, magnetic field effects on the phase diagram in Ho \cite{Rodrigues} and a few experiments \cite{Karhu2011,Karhu2012}. Helical magnets present potential applications in spintronics with predictions of spin-dependent electron transport in these magnetic materials \cite{Heurich,Wessely,Jonietz}. 
%
%

We consider a thin film of BCC lattice of $N_z$ layers, with two symmetrical surfaces perpendicular to
the $c$-axis, for simplicity.  The exchange Hamiltonian reads

\begin{equation}\label{eqn:hamil}
\mathcal H=\mathcal H_e+\mathcal H_a
\end{equation}
where the isotropic exchange part is given by
\begin{equation}
\mathcal H_e=-\sum_{\left<i,j\right>}J_{i,j}\mathbf S_i\cdot\mathbf
S_j  \label{eqn:hamil1}
\end{equation}
$J_{i,j}$ being the interaction between two quantum Heisenberg spins $\mathbf S_i$ and $\mathbf S_j$ occupying the lattice sites $i$ and $j$.
The anisotropic part is chosen as
\begin{equation}\label{eqn:hamil2}
\mathcal H_a= -\sum_{<i,j>} I_{i,j}S^z_iS^z_j\cos\theta_{ij}
\end{equation}
where $I_{i,j}$ is the anisotropic interaction along the in-plane local spin-quantization axes $z$ of $\mathbf S_i$
and $\mathbf S_j$,  supposed to be positive, small compared to $J_1$, and limited to NN on the $c$-axis.
Let us mention that according to the theorem of Mermin and Wagner \cite{Mermin}
 continuous isotropic spin models such as XY and Heisenberg spins
do not have long-range ordering at finite temperatures in two dimensions. Since we are dealing
with the Heisenberg model in a thin film, it is useful to add an anisotropic interaction
to reinforce a long-range ordering at finite temperatures.

To generate a bulk helimagnetic structure, the simplest way is to take a ferromagnetic interaction between NNs, say $J_1$ ($>0$),
and an antiferromagnetic interaction between NNNs,  $J_2<0$.  It is obvious that if $|J_2|$ is smaller than a critical value $|J_2^c|$,
the classical GS spin configuration is ferromagnetic \cite{Rastelli,Diep89}.
Since our purpose is to investigate the helimagnetic structure near the surface and surface effects, let us consider the case of a
helimagnetic structure only in the $c$-direction perpendicular to the film surface. In such a case, we assume a non-zero $J_2$ only on the $c$-axis.
This assumption simplifies formulas but does not change the physics of the problem since including the uniform helical angles in two other
directions parallel to the surface will not introduce additional surface effects.  Note that the bulk case of the above
quantum spin model has been studied by the Green's function method \cite{Quartu1998}.

To calculate the classical GS surface spin configuration, we write down the expression of
the energy of spins along the $c$-axis, starting from the surface:
\begin{eqnarray}
E&=& -Z_1 J_1 \cos (\theta_1-\theta_2)-Z_1 J_1 [\cos (\theta_2-\theta_1)\nonumber\\
&&+ \cos (\theta_2-\theta_3)]+...\nonumber\\
&&-J_2 \cos (\theta_1-\theta_3)-J_2 \cos (\theta_2-\theta_4)\nonumber\\
&&-J_2[\cos (\theta_3-\theta_1)+ \cos (\theta_3-\theta_5)]+...\label{EC}
\end{eqnarray}
where $Z_1=4$ is the number of NNs in a neighboring layer, $\theta_i$ denotes the angle of a spin in the $i$-th layer
made with the Cartesian $x$ axis of the layer. The interaction energy between two NN spins in the two adjacent layers $i$ and $j$
depends only on the difference $\alpha_{i}\equiv \theta_i-\theta_{i+ 1}$. The GS configuration corresponds to the minimum of $E$.
We have to solve the set of equations:
\begin{equation}
\frac{\partial E}{\partial \alpha_i}=0, \ \ \ \mbox{for}\ \ i=1,N_z-1
\end{equation}
Explicitly, we have
\begin{eqnarray}
\frac{\partial E}{\partial \alpha_1}&=&8J_1\sin \alpha_1+2J_2\sin (\alpha_1+\alpha_2)=0\label{A1}\\
\frac{\partial E}{\partial \alpha_2}&=&8J_1\sin \alpha_2+2J_2\sin(\alpha_1+\alpha_2)\nonumber\\
&&+2J_2\sin(\alpha_2+\alpha_3)=0\label{A2}\\
\frac{\partial E}{\partial \alpha_3}&=&8J_1\sin \alpha_3+2J_2\sin(\alpha_2+\alpha_3)\nonumber\\
&&+2J_2\sin(\alpha_3+\alpha_4)=0\label{A3}\\
\frac{\partial E}{\partial \alpha_4}&=&... \nonumber
\end{eqnarray}
where we have expressed the angle between two NNNs as follows: $\theta_1-\theta_{3}=\theta_1-\theta_{2}+ \theta_2-\theta_{3}=\alpha_1+\alpha_2$ etc.

Solving these equations, we obtain the results shown in Table \ref{table}.  
\begin{center}
\begin{table}
\begin{tabular}{|l|c|c|c|c|r|}
\hline
$J_2/J_1$ & $\cos \theta_{1,2}$ & $\cos\theta_{2,3}$ & $\cos\theta_{3,4}$ & $\cos\theta_{4,5}$ & $\alpha$(bulk)  \\
\hline
&&&&&\\
-1.2  &  0.985($9.79^\circ$) &  0.908($24.73^\circ$)       &    0.855($31.15^\circ$)    &   0.843($32.54^\circ$)  &   $33.56^\circ$    \\
-1.4 &  0.955($17.07^\circ$)&  0.767($39.92^\circ$)  &  0.716($44.28^\circ$)    &   0.714($44.41^\circ$)   &  $44.42^\circ$     \\
-1.6 & 0.924($22.52^\circ$) &  0.633($50.73^\circ$) & 0.624($51.38^\circ$)  &  0.625($51.30^\circ$)   &  $51.32^\circ$        \\
-1.8     &  0.894($26.66^\circ$)  &  0.514($59.04^\circ$)  & 0.564($55.66^\circ$)   &  0.552($56.48^\circ$)  &  $56.25^\circ$    \\
-2.0       &  0.867($29.84^\circ$)  &  0.411($65.76^\circ$) &  0.525($58.31^\circ$)   &  0.487($60.85^\circ$)  & $60^\circ$   \\
&&&&&\\
\hline
\end{tabular}
\caption{ Values of $\cos \theta_{n,n+1}=\alpha_n$ between two adjacent layers are shown for various
values of $J_2/J_1$ of a film of 8 layers. Only angles of the first half are shown: other angles are, by symmetry,
$\cos\theta_{7,8}$=$\cos\theta_{1,2}$, $\cos\theta_{6,7}$=$\cos\theta_{2,3}$, $\cos\theta_{5,6}$=$\cos\theta_{3,4}$. The last column shows the value of the angle in the bulk case (infinite thickness).} \label{table} 
\end{table}
\end{center}

We see that  strong angle variations  occur near the surface with oscillations for strong $J_2$. This is called "surface spin reconstruction".  Note that the angles at the film center are close to the bulk value $\alpha$ (last column), meaning that the surface reconstruction affects just a few atomic layers even for films thicker that $N_z=8$. This helical stability has been experimentally observed in holmium films \cite{Leiner}.

We have numerically performed the numerical steepest descent method described in Ref. \cite{Ngo2006a} and we find the analytical result shown in Table I.  

\begin{figure}[htb]
\centering
\includegraphics[width=3.2 in]{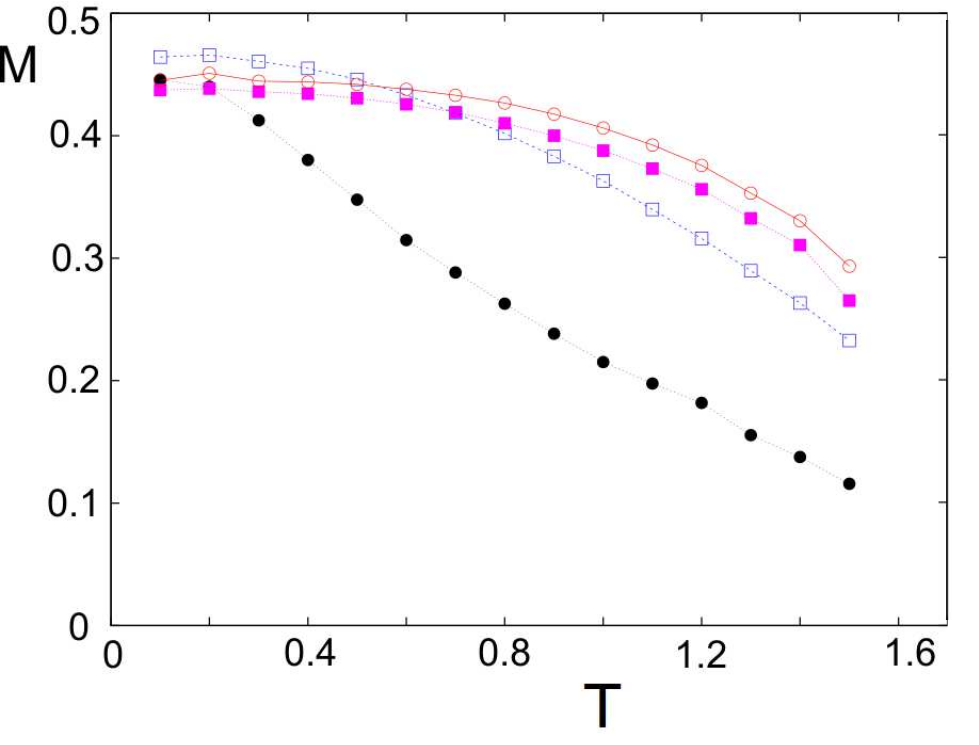}
\caption{ Layer magnetizations versus $T$ for the surface interaction $J_1^s=0.3$  with $J_2/J_1=-2$, $d=0.1$ and $N_z=16$. Black circles, blue void squares, magenta squares and red void circles are for first, second, third and fourth layers, respectively.}\label{surf03}
\end{figure}

In Ref. \cite{DiepHeli} we have used the Green's function method devised for non-collinear spin configurations \cite{QuartuDiep}. The calculations are lengthy, the reader is referred to those papers for the detailed of the Geen's function theory applied to non-collinear spin configurations. The  obtained spin-wave spectrum obtained from this theory allows us  to calculate the quantum fluctuations at $T=0$, the magnetization layer by layer versus $T$, and the transition temperature.  An example is shown in Fig. \ref{surf03} for the layer magnetizations as functions of $T$. Note that the surface magnetization is very low at finite $T$ with respect to interior layers.  Note also that the layer magnetizations are smaller than the spin length $S=0.5$ at $T=0$. This is due to the zero-point quantum fluctuations mentioned above: this phenomenon is called "zero-point spin contraction" found in quantum antiferromagnets \cite{DiepTM}.

Another  example of surface spin configuration different from the bulk one is shown in Example 2 below.

\subsection{Surface phase transitions}
In the presence of a surface, interaction parameters between spins on the surface and spins on the beneath layer may be different from the bulk interaction parameters. As a consequence, when the temperature $T$ increases, one may observe a phase transition at the surface distinct from that of the bulk.   We show below some examples of this phenomenon:

{\bf Example 2} We consider a thin film made up by stacking $N_z$
planes of triangular lattice of $N\times N$ lattice sites.

The Hamiltonian is given by
\begin{equation}
\mathcal H=-\sum_{\left<i,j\right>}J_{i,j}\mathbf S_i\cdot\mathbf
S_j -\sum_{<i,j>} I_{i,j}S_i^z S_j^z  \label{eqn:hamil1}
\end{equation}
where $\mathbf S_i$ is the quantum Heisenberg spin of magnitude $S=0.5$ at the lattice site
$i$, $\sum_{\left<i,j\right>}$ indicates the sum over the NN spin
pairs  $\mathbf S_i$ and $\mathbf S_j$.  The last term, which will
be taken to be very small,  is needed to make the film with a
small finite thickness to have a clear phase transition at a finite
temperature in the case where all exchange interactions $J_{i,j}$
are ferromagnetic. This guarantees the existence of a phase
transition at finite temperature, since it is known that a
strictly 2D system with an isotropic non-Ising spin
model (XY or Heisenberg model) does not have long-range ordering
at finite temperatures \cite{Mermin}.

To be general, we take the interaction between two NN spins lying on the surface spins to be $J_s$.
Interaction between layers and interaction between NN in interior
layers are supposed to be ferromagnetic and all equal to $J=1$ for
simplicity. The two surfaces of the film are frustrated if $J_s$
is antiferromagnetic ($J_s<0$). Note that if the surface is alone, we have the planar $120^{\circ}$ structure.

\begin{figure}[ht]
\centering
\includegraphics[width=3.2 in]{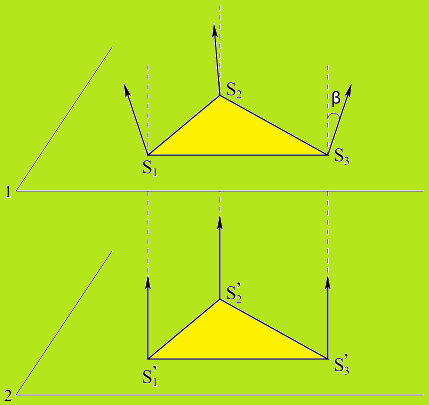}
\caption{Non-collinear surface spin configuration. Angles between spins on
layer $1$ are all equal (noted $\alpha$), while angles between
vertical spins are $\beta$.} \label{fig:gsstruct}
\end{figure}

With the above model we have shown \cite{Ngo2006a} that the ground state is non-collinear near the surface: the three spins on a triangle on the surface form  an 'umbrella' with an angle $\alpha$ between them and an angle
$\beta$ between a surface spin and its beneath neighbor (see Fig.
\ref{fig:gsstruct}). This non-planar structure is due to the
interaction of the spins on the beneath layer, just like an
external applied field in the $z$ direction. Of course, when $J_s$
is larger than $J_s^c$ one has the collinear ferromagnetic GS as
expected: the frustration is not strong enough to resist the
ferromagnetic interaction from the beneath layer. 

\begin{figure}[htb!]
\centering
\includegraphics[width=12cm]{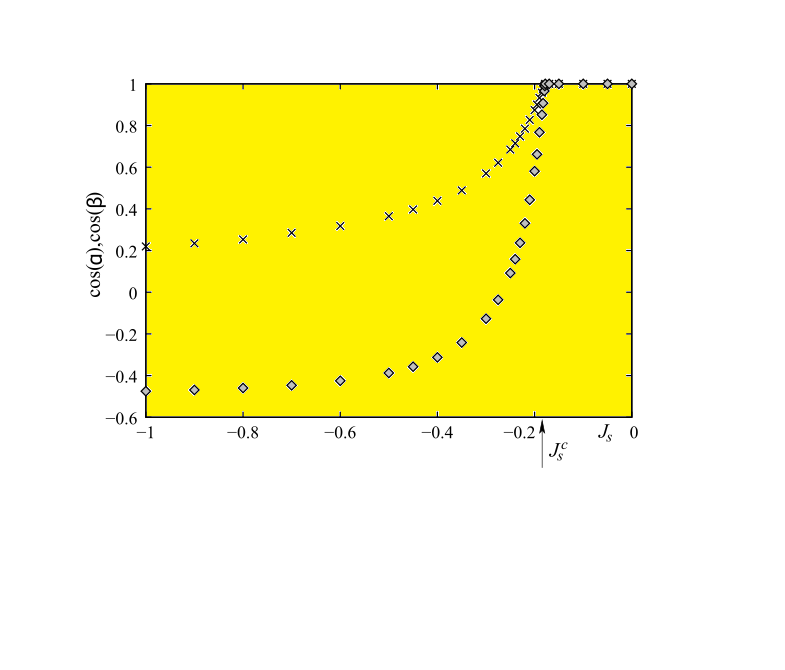}
\vspace{-3cm} 
\caption{Non-collinear surface spin configuration. Angles between spins on
layer $1$ are all equal (noted $\alpha$), while angles between
vertical spins are $\beta$.  The arrow indicates the critical value $J_s^c \approx -0.1889 J $}. \label{fig:gscos}
\end{figure}

We show in Fig. \ref{fig:gscos} $\cos(\alpha)$ and $\cos(\beta)$
as functions of $J_s$. The critical value $J_s^c$ is found between
-0.18 and -0.19.  This value can be calculated analytically by
assuming the 'umbrella structure' \cite{Ngo2006a} shown in Fig. \ref{fig:gsstruct}, we obtain

\begin{equation}
\cos\beta = -\frac{J+I}{9J_s+6I_s}. \label{eqn:GSsolu}
\end{equation}
For  given values of $I_s$ and $I$, we see that the solution
(\ref{eqn:GSsolu}) exists for $J_s \leq J_s^c$ where the critical
value $J_s^c$ is determined by $-1\leq \cos\beta \leq 1$. For
$I=-I_s=0.1$, $J_s^c \approx -0.1889 J $.

\begin{figure}[hbt!]
\centering
\includegraphics[width=12cm]{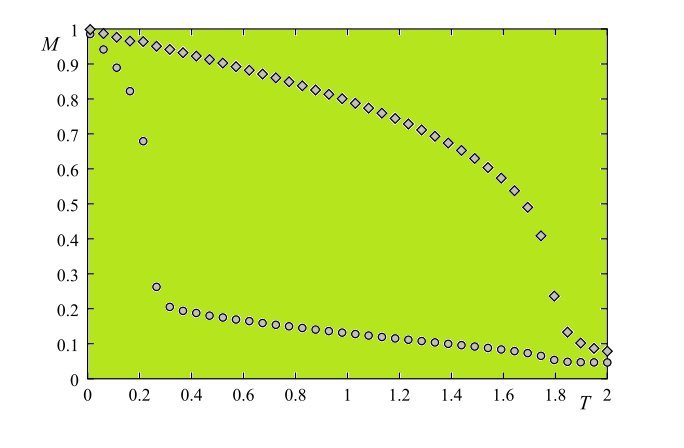}
\caption{Magnetizations of layer 1 (circles) and layer 2
(diamonds) versus temperature $T$ in unit of $J/k_B$ for
$J_s=-0.5$ with $I=-I_s=0.1$.} \label{fig:HSn05Ms}
\end{figure}

Using MC simulations, we showed that in the case of highly frustrated surface, there is a surface phase transition occurring at low $T$ as seen in Fig. \ref{fig:HSn05Ms}: the surface magnetization ($M_1$) falls sharply at $T_1\simeq 0.25J/k_B$ while the magnetization of the beneath layer ($M_2$) falls at $T_2\simeq 1.8 J/k_B$. Note that between $T_1$ and $T_2$, $M_1$ is not zero because of $M_2$ acting on it as an external field.

To conclude this section, we note that in Ref. \cite{Ngo2006a}, we have also studied the quantum counterpart of the model (\ref{eqn:hamil1}) using the Green's function. The results show quantum fluctuations of the layer magnetizations at low $T$ but we obtain as expected the main features as found above at high $T$ in Fig. \ref{fig:HSn05Ms} .

\section{Criticality of thin films}\label {critical}

The theory of criticality of phase transitions was demonstrated in the case of infinite crystals.  The beauty of the theory resides in the fact that, despite an infinite number of systems, the phase transitions belong to a small number of "universality class".  The university class depends on a few numbers of parameters: the interaction nature (short range, long range, ...), the symmetry of the order parameter (spins of Ising, XY, Heisenberg, Potts,...) and the space dimension (2D, 3D,...). Each universality class is  defined by 6 critical exponents.
When the transition is of second order, one can define in the
vicinity of $T_c$ the following critical exponents \index{critical
exponents}

\begin{eqnarray}
C_V&=&A\left|\frac {T-T_c}{T_c}\right|^{-\alpha}\label{transit4}\\
\overline M&=&B\left [\frac {T_c-T}{T_c}\right ]^{\beta}\label{transit5}\\
\chi&=&C\left|\frac {T-T_c}{T_c}\right|^{-\gamma}\label{transit6}\\
\xi &\propto &\left [\frac {T-T_c}{T_c}\right ]^{-\nu}\label{transit7}\\
\overline M&=&H^{1/\delta} \label{transit8}
\end{eqnarray}

Note that the same $\alpha$ is defined for $T>T_c$ and $T<T_c$ but the
coefficient $A$ is different for each side of $T_c$.  This is also
the case for $\gamma$. However, $\beta$ is defined only for $T<T_c$ because $\overline M=0$ for $T\geq T_c$.
The definition of $\delta$ is valid only at $T=T_c$ when the system
is under an applied magnetic field of amplitude $H$.  Finally, at
$T=T_c$ one defines exponent $\eta$ of the correlation function by
\begin{equation}\label{transit9}
G( r) \propto \frac{1}{r^{d-2+\eta}}
\end{equation}

There is another exponent called ``dynamic exponent" $z$
defined via the relaxation time $\tau $ of
the spin system for  $T\geq T_c$:

\begin{equation}\label{transit10}
\tau \propto \xi^z \propto \left [\frac{1} {T-T_c}\right ]^{z\nu}
\end{equation}

There are thus six critical exponents $\alpha$, $\beta$, $\gamma$, $\delta$, $\nu $
and $\eta $. However, there are four relations between them: \cite{Wilson,Amit,Cardy,Zinn}

\begin{eqnarray}
\alpha +2\beta +\gamma&=&2 \label{transit51} \\
\alpha+\beta (1+\delta)&=&2 \label{transit52}\\
\alpha&=&2-d\nu \label{transit55}\\
\gamma&=&\nu (2-\eta) \label{transit56}
\end{eqnarray}
Therefore, there are only two of them are to be computed. The last two relations are called "hyperscaling relations".  

Note that when a phase transition is of first order, the first derivatives of the free energy such as the internal energy and the order parameter (magnetization, for example) are discontinuous at the phase transition, one cannot defined the critical exponents given above. The phase transition is not "critical". 

Now, if we have a thin film where the thickness is finite, does the theory of criticality apply? The answer is difficult as we see by the work presented below. Also, when a transition is of first order in 3D, does it remain so in a film with a finite thickness? We show examples of the two mentioned cases below.

\subsection{ Critical exponents of thin films:}

In Ref. \cite{PhamSS2009}, we have studied  the question whether or not the phase transition known in
the bulk state changes its nature when the system is made as a thin film.
We have considered the case of a bulk second-order transition. For that purpose, we studied the case of an Ising ferromagnet. 
The criticality of a film with uniform
interaction has been studied by Capehart and Fisher
as a function of the film thickness using a scaling
analysis\cite{Fisher} and by MC simulations.\cite{Schilbe,Caselle}
The results by Capehart and Fisher indicated that as long as the
film thickness is finite the phase transition is strictly  that of
the 2D Ising universality class.  However, they showed that at a
temperature away from the transition temperature $T_c(N_z)$, the
system can behave as a 3D one when the spin-spin correlation
length $\xi(T)$ is much smaller than the film thickness, i. e.
$\xi(T)/N_z\ll 1$. As $T$ gets very close to $T_c(N_z)$, 
$\xi(T)/N_z \rightarrow 1$, the system undergoes a crossover to 2D
criticality.  We will return to this work for comparison with our
results shown below.

\begin{figure}
\includegraphics[width=8cm]{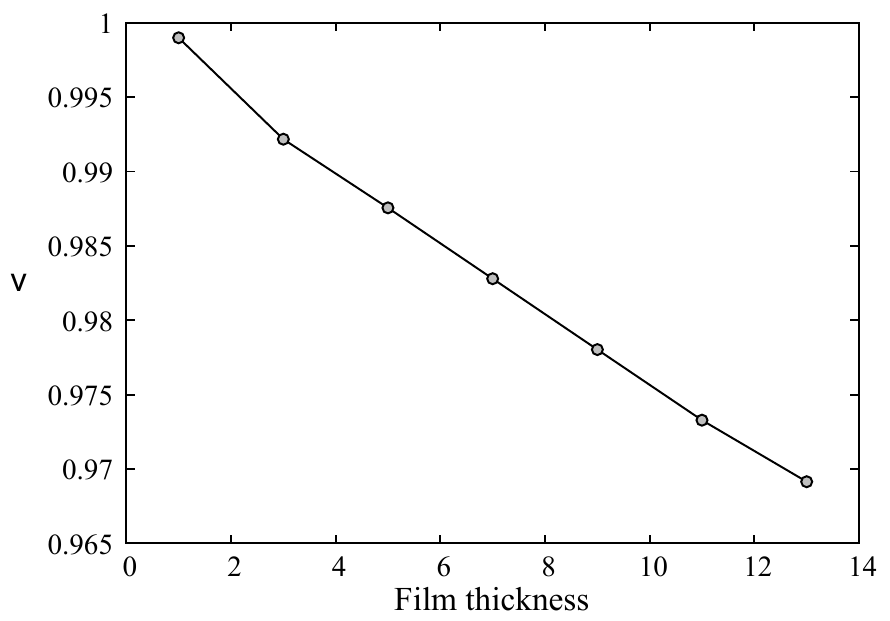} 
\caption{Effective
exponent $\nu$ versus $N_z$.} \label{fig:NUZ}
\end{figure}

We have used the high-performance multiple histogram technique which is known to reproduce with very
high accuracy the critical exponents of second order phase
transitions.\cite{Ferrenberg1,Ferrenberg2,Ferrenberg3,Bunker}
The reader is referred to Ref. \cite{PhamSS2009} for technical details, specially on the finite-size scaling relations. We have shown that under a thin film shape, i.e. with a finite thickness, the transition
shows "effective critical exponents" whose values are between 2D and 3D universality classes.
As an example, we show the exponent $\nu$ as a function of the film thickness in Fig. \ref{fig:NUZ} (other exponents can be found in Ref. \cite{PhamSS2009}).

The deviation of $\nu$ from the 2D value when $N_z$ increases is
due, as discussed earlier, to the crossover to 3D.  Other exponents suffer the same deviations. Note that
these results are in excellent agreement with
the exact results $\nu_{2D}=1$ (and $\gamma_{2D}=1.75$ not shown here).  The very
high precision of our method is thus verified in the rather modest
range of the system sizes $L=20-80$ used in the present work. Note
that the result of Ref.\cite{Schilbe} gave $\nu=0.96\pm0.05$
for $N_z=1$ which is very far from the exact value.

\begin{figure}
\includegraphics[width=8cm]{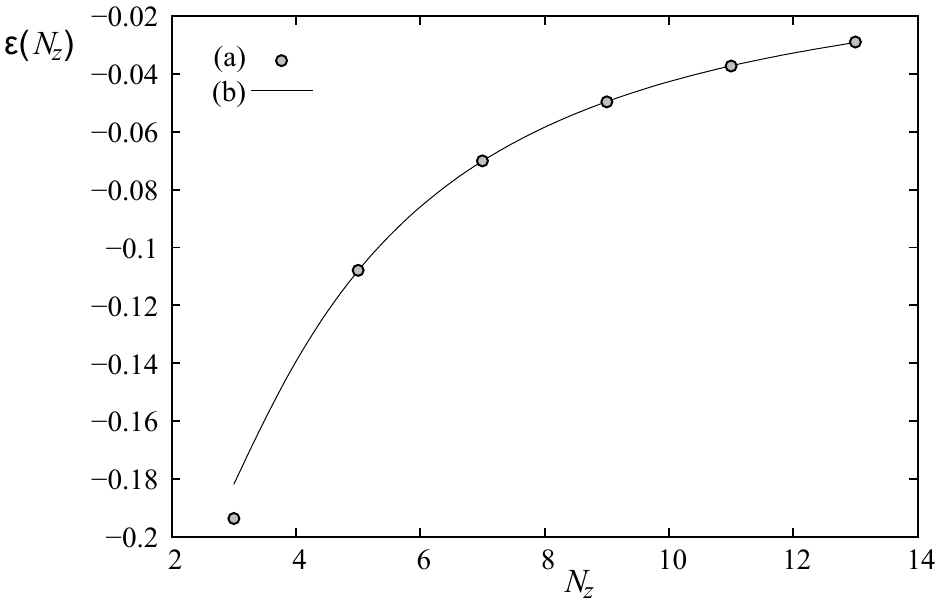} 
\caption{Critical
temperature at infinite $L$ as a function of the film thickness.
Points are MC results, continuous line is the prediction of
Capehart and Fisher. The agreement is excellent. }
\label{TCINF}
\end{figure}

Let us show the prediction of Capehart and Fisher\cite{Fisher} on
the critical temperature as a function of $N_z$.
 Defining the critical-point shift as
\begin{equation}
\varepsilon(N_z)=\left[ T_c(L=\infty,N_z)-T_c(3D)\right]/T_c(3D)
\end{equation}
they showed that

\begin{equation}\label{CF}
\varepsilon(N_z)\approx \frac{b}{N_z^{1/\nu}}[1+a/N_z]
\end{equation}
where $\nu=0.6289$ (3D value). Using  $T_c(3D)=4.51$,  we fit the
above formula with $T_c(L=\infty,N_z)$ obtained from our MC simulations (see  Ref. \cite{PhamSS2009}), 
 we should find,
as long as the thickness is finite, the 2D universality class. This has been done, the scaling of Capehart and Fisher is validated by our results: the MC
results and the fitted curve are shown in Fig. \ref{TCINF}. Note
that if we do not use the correction factor $[1+a/N_z] $, the fit
is not good for small $N_z$.  The prediction of Capehart and
Fisher is thus very well verified.

\subsection{Cross-over from first- to second-order transition in thin films:}

We have shown in Ref. \cite{PhamCrossover} the crossover of the phase transition from first to second order in
the frustrated Ising FCC AF film.  This crossover occurs when the film thickness $N_z=2$
is smaller than a value between 2 and 4 FCC lattice cells.  These results are obtained with the highly performing Wang-Landau flat
histogram technique which allows to determine a weak first-order transition with efficiency. \cite{WL1,WL2,brown,Schulz,Malakis}
This algorithm is very efficient for classical statistical models.
The algorithm uses a random walk in energy space in order to obtained an accurate
estimate for the density of states $\rho(E)$ which is defined as the number of spin
configurations for any given energy $E$. 

For $N_z=2$, we found that in a range of temperature the surface spins stay ordered  while interior spins are disordered. We interpret this as an effect of the frustration reduction: due to the lack of neighbors, the surface spins are less  frustrated
than the interior spins. As a consequence, interior spins are disordered at a lower temperature.  This has been verified by the Green's function calculation.\cite{PhamCrossover}

The second-order transition for $N_z=2$ is governed by the surface disordering and is characterized by  critical exponents whose values are deviated from those of the 2D Ising universality class. We believe that this deviation  results from the effect of the disordered interior spins which act as "correlated" random fields on the surface spins.  We do not know if the critical exponents found here belong to a new universality class or they are just "effective critical exponents" which one could scale in some way or another to bring into the 2D Ising universality class.  Anyway, these exponents seem to obey a weak universality.\cite{Suzuki}  An answer to this question is still desirable.

\section{Spin waves in a monolayer and thin films with Dzyaloshinskii-Moriya interaction}\label{DMSW}
The Dzyaloshinskii-Moriya (DM) interaction was historically introduced to explain the weak ferromagnetism which was experimentally seen in antiferromagnetic Mn compounds \cite{Dzyaloshinskii,Moriya}.   The phenomenological Landau-Ginzburg model introduced by I. Dzyaloshinskii \cite{Dzyaloshinskii} was microscopically derived by T. Moriya \cite{Moriya}. The interaction between two spins $\mathbf S_i$ and $\mathbf S_j$ is written as

\begin{equation}\label{eq1}
 \mathbf D_{i,j}\cdot \mathbf S_i\wedge\mathbf S_j
\end{equation}
where $\mathbf D_{i,j}$ is a vector which results from the displacement of non magnetic ions located between $\mathbf S_i$ and $\mathbf S_j$,  for example in Mn-O-Mn bonds. The direction of  $\mathbf D_{i,j}$ depends on the symmetry of the displacement \cite{Moriya}.  The DM interaction is antisymmetric with respect to the inversion symmetry.

There has been a large number of investigations on the effect of the DM interaction in various materials, both experimentally and theoretically for weak ferromagnetism in perovskite compounds (see references cited in Refs. \onlinecite{Sergienko,Ederer}, for example). However, the interest in the DM interaction goes beyond the weak ferromagnetism: for example, it has been recently shown in various works that the DM interaction is at the origin of topological skyrmions which will be shown in the next section.

Hereafter, we show a simple system of Heisenberg spins interacting with each orthe via a DM interaction in addition to a ferromagntic exchange interaction. Consider first the case of a monolayer opf square lattice. 
The Hamiltonian is given by

\begin{eqnarray}
\mathcal H&=&\mathcal H_e+\mathcal H_{DM}\label{eqn:hamil1}\\
\mathcal H_e&=&-\sum_{\left<i,j\right>}J_{i,j}\mathbf S_i\cdot\mathbf S_j\label{eqn:hamil2}\\
\mathcal H_{DM}&=&\sum_{\left<i,j\right>}\mathbf D_{i,j}\cdot \mathbf S_i\wedge\mathbf S_j
 \label{eqn:hamil3}
\end{eqnarray}
where $J_{i,j}$ and $\mathbf D_{i,j}$ are the exchange and DM interactions, respectively,
between two Heisenberg spins $\mathbf S_i$ and $\mathbf S_j$ of magnitude $S=1/2$
occupying the lattice sites $i$ and $j$.

The SC lattice can support the DM interaction in the absence of the
inversion symmetry \cite{Moriya,Maleyev,Muhlbauer} as in MnSi.
The absence of inversion symmetry can be also achieved by the positions of non-magnetic ions between magnetic ions.
The vector $\mathbf D$ between two magnetic ions is defined as $\mathbf D_{i,j}= A\mathbf r_i \wedge \mathbf r_j$ where $\mathbf r_i$ is the vector connecting the non-magnetic ion to the spin $\mathbf S_i$ and $\mathbf r_j$ is that to the spin $\mathbf S_j$, $A$ being a constant.  One sees that $\mathbf D_{i,j}$ is perpendicular to the plane formed by $\mathbf r_i$ and $\mathbf r_j$.

For simplicity, let us consider the case where the in-plane  exchange interactions between NN
is ferromagnetic and denoted by $J$.
The vector $\mathbf D_{i,j}$  is chosen by supposing the situation where non-magnetic ions are positioned in the plane. With this choice,
$\mathbf D_{i,j}$ is perpendicular to the monolayer plane (let's name it $xz$ plane) and our model gives rise to a planar spin configuration, namely the spins stay in $xz$ plane. There are no situations where the spins are out of plane. This simplifies our calculation.
The DM interaction is supposed to be between NN in the plane with a constant $D$.

\begin{figure}[ht!]
\centering
\includegraphics[width=8cm,angle=0]{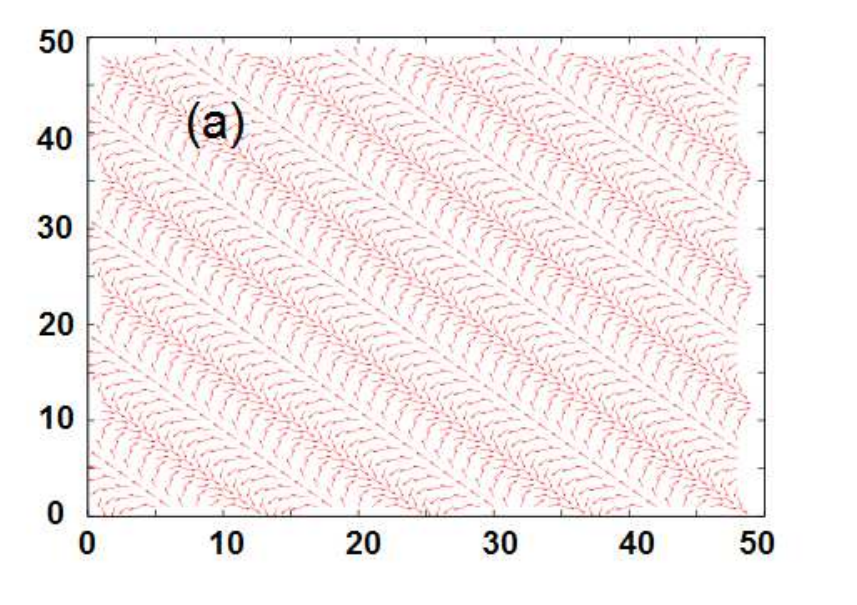}
\includegraphics[width=8cm,angle=0]{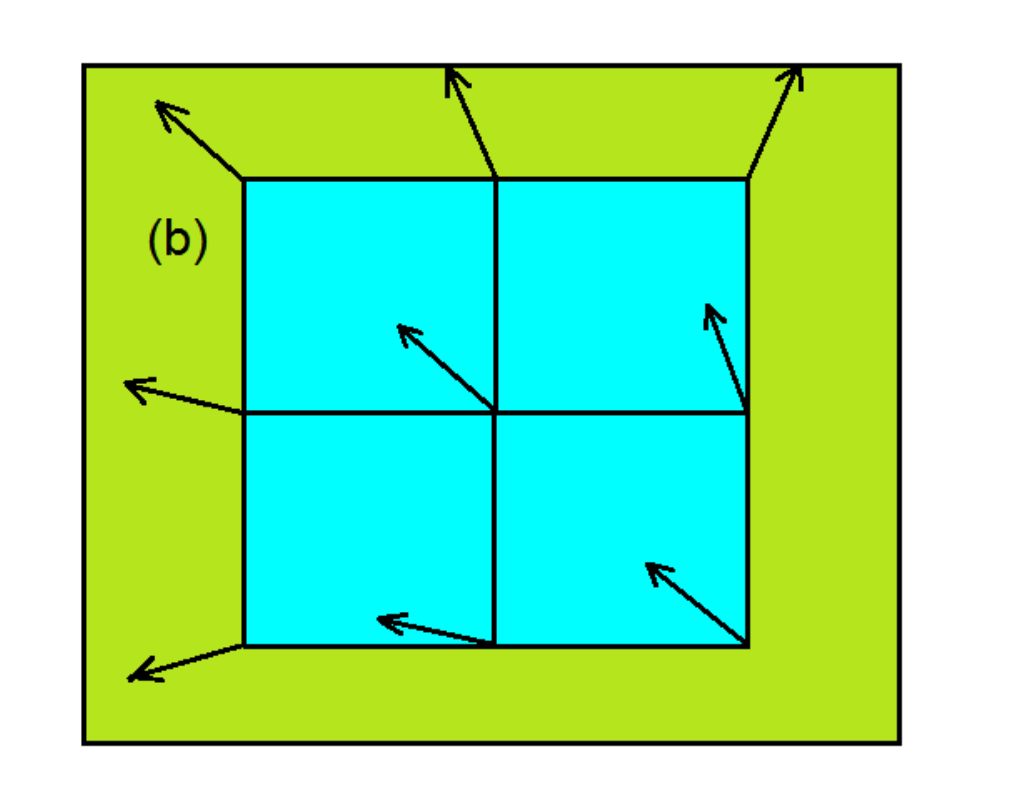}
\caption{(a) The ground state configuration on the $xz$ plane. The figure shows the case where
$\theta=\pi/6$ ($D=-0.577)$ along the $x$ and $z$ axes, with $J=1$, obtained by using
the steepest descent method ; (b) a zoom is shown around a spin with its nearest neighbors in the $xz$ plane. \label{ffig1}}
\end{figure}

Due to the competition between the exchange $J$ term which favors the collinear configuration, and the
DM term which favors the perpendicular one, we expect that the spin $\mathbf S_i$ makes
an angle $\theta_{i,j}$ with its neighbor $\mathbf S_j$. Considering the lattice symmetry  and the uniform interactions, we deduce that the energy of the spin $\mathbf S_i$ is written as
\begin{equation}
E_{i}=-4JS^2\cos\theta+ 4DS^2\sin\theta
\end{equation}
where $\theta=|\theta_{i,j}|$ and care has been taken on the signs of $\sin \theta_{i,j}$ 
when counting NN, namely
two opposite NN have opposite signs for the sinus term.  The minimization
of $E_i$ yields
\begin{equation}
\frac{dE_{i}}{d\theta}=0\  \ \Rightarrow \  \  -\frac{D}{J}=\tan \theta
\  \ \Rightarrow \  \  \theta=\arctan (-\frac{D}{J})\label{gsangle}
\end{equation}
The value of $\theta$ for a given $\frac{D}{J}$ is precisely what obtained by
the steepest descent method \cite{Diep2017}.  We show in Fig. \ref{ffig1} the relative orientation of the two NN spins in the plane $xz$ obtained by the mentioned steepest descent method.

\begin{figure}[h!]
\center
\includegraphics[width=6cm]{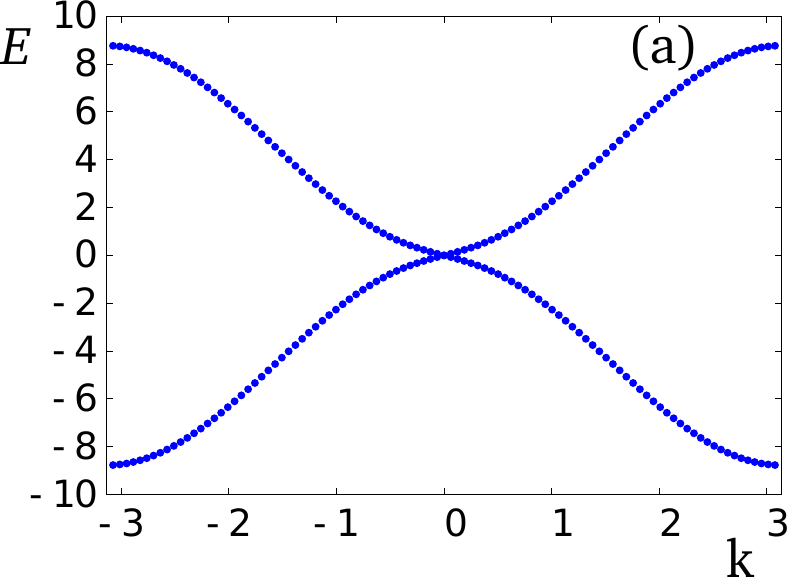}
\includegraphics[width=6cm]{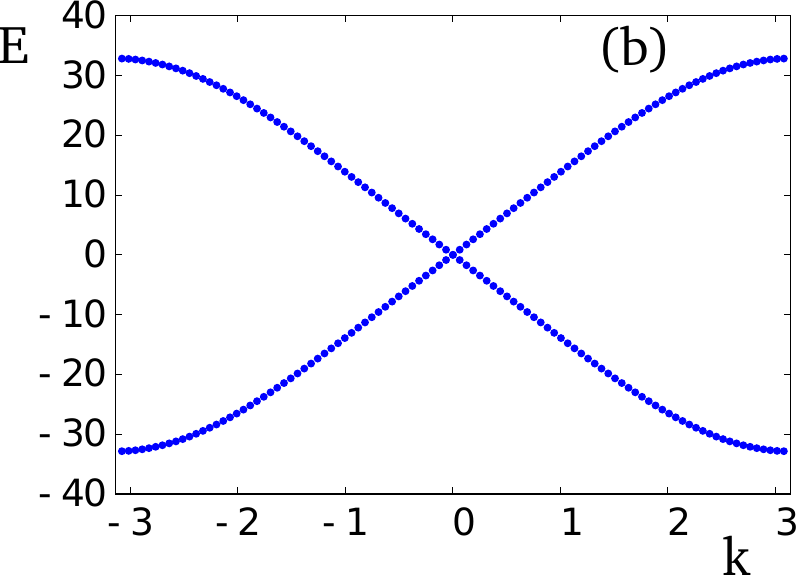}
\caption{Spin-wave spectrum $E (k)$ versus $k\equiv k_x=k_z$ for (a) $\theta=0.524$ radian and (b) $\theta=1.393$ in two dimensions at $T=0.1$.
Positive and negative branches correspond to
right and left precessions. 
\label{ffig3}}
\end{figure}

To calculate the spin-wave spectrum and various thermodynamic quantities such as the magnetization versus temperature $T$, we use the Green's fuction technique devised for non-collinear spin configurations  \cite{QuartuDiep}. The reader is referred to the paper \cite{Diep2017} for the details of the calculation.  We show in Fig. \ref{ffig3} the SW spectrum calculated  for $\theta=30$ degrees ($\pi/6$ radian) and $80$ degrees (1.396 radian). The spectrum is symmetric for positive and negative wave vectors and for left
and right precessions. Note that for small $\theta$ (i. e. small $D$) $E$ is proportional to $k^2$ at low $k$ (cf. Fig. \ref{ffig3}a), as in ferromagnets. However, as $\theta$ increases, we observe that $E$ becomes linear in $k$ at low $k$ as seen in Fig. \ref{ffig3}b. This is similar to antiferromagnets.  The change of behavior is progressive with increasing $\theta$, we do not observe a sudden transition from $k^2$ to $k$ behavior.  This feature is also observed in three dimensions (3D) (see \cite{Diep2017}).

\begin{figure}[ht!]
\centering
\includegraphics[width=7cm,angle=0]{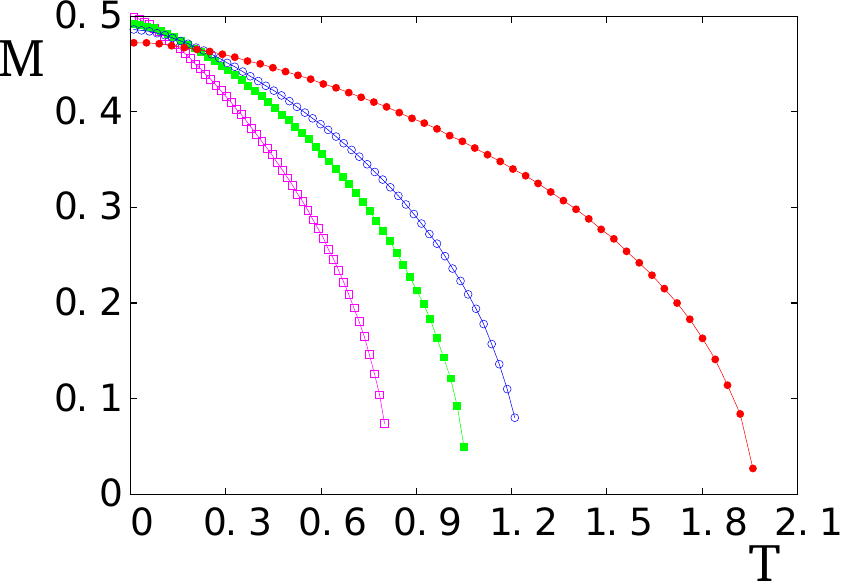}
\caption{Magnetizations $M$ versus temperature $T$ for a monolayer (2D) $\theta=0.175$ (radian), $\theta=0.524$, $\theta=0.698$,  $\theta=1.047$ (void magenta squares,  green filled squares, blue void circles and filled red circles, respectively). 
\label{ffig4}}
\end{figure}

We show in Fig. \ref{ffig4} the magnetization versus $T$ for various $\theta$. Note that for strong values of 
$\theta$ (strong angles) the magnetization is less than 1/2 at $T=0$ because of the so-called "zero-point spin contraction" due to quantum fluctuations at $T==0$ as seen in antiferromagnets (see \cite{DiepTM}). 

The cases of thin films and 3D have also been calculated in Ref. \cite{Diep2017}.

To conclude this section, we would like to emphasize that the Green's function technique is, to our knowledge, the only one which can calculate the spin-wave spectrum and the magnetization versus $T$ for  non-collinear spin configurations.

\section{Skyrmions in a monolayer}\label{Skyrmions}
A skyrmion is a vortex-like spin configuration which has a double chirality.  It has a variety of origin: we can mention spin systems with Dzyaloshinskii-Moriya (DM) interaction \cite{Dzyaloshinskii,Moriya}, highly frustrated systems,\cite{Hayami,Okubo,Rosales} systems with combined competing interactions,... Due to the robustness of skyrmions and their small sizes (a dozen of atoms), there are many potential applications in spintronics. \cite{Fert2013}  This explains an enormous number of publications on skyrmions every single day.  We shall show some examples in the following.

Let us mention the important papers appeared since 2003 \cite{Bogdanov2003}, from Ref. \cite{Leonov} to Ref. \cite{Sahbi2018}.  The reader is invited to look at the titles of these references to have an idea on what has  been done. Surely, I miss a lot of important papers.   

Hereafter, we consider for simplicity the two-dimensional case where the spins are on a square lattice in the $xy$ plane of linear dimension $N$. We are interested in the stability of the skyrmion crystal generated in a system of spins interacting with each other via a DM interaction and a symmetric isotropic Heisenberg exchange interaction in an applied field perpendicular to the $xy$ plane. All interactions are limited to NN. The results below have been published in Ref. \cite{Sahbi2018}.
The full Hamiltonian is given by

\begin{eqnarray}
\mathcal{H}&=&-J \sum_{\langle ij \rangle} \mathbf{S_i} \cdot \mathbf{S_j} +D \sum_i \mathbf{S_i} \wedge (\mathbf{S}_{i+x}  +\mathbf{S}_{i+y} )\nonumber\\
&&-H \sum_i S_i^z
\end{eqnarray}
where the DM interaction and the exchange interaction are taken between NN on both $x$ and $y$ directions.  Rewriting it it in a convenient form, we have
\begin{eqnarray}
\mathcal{H}&=&-J\sum_{\langle ij \rangle} \mathbf{S_i} . \mathbf{S_j} +D \sum_i [S_i^y S_{i+x}^z-S_i^z S_{i+x}^y-S_i^x S_{i+y}^z\nonumber\\
&&+S_i^z S_{i+y}^x]- H\sum_i S_i ^z\nonumber\\
&=&-J \sum_{\langle ij \rangle} \mathbf{S_i} . \mathbf{S_j} +D \sum_i [S_i^y (S_{i+x}^z-S_{i-x}^z)\nonumber\\
&&-S_i^z (S_{i+x}^y-S_{i-x}^y)-S_i^x (S_{i+y}^z-S_{i-y}^z)\nonumber\\
&&+S_i^z( S_{i+y}^x- S_{i-y}^x)]
- H \sum_i S_i^z
\end{eqnarray}

For the $i$-th spin, one can write
\begin{equation}\label{LF}
\mathcal{H}_i= -S_i^x H_i^x - S_i^y H_i^y - S_i^z H_i^z
\end{equation}
where the local-field components are given by
\begin{eqnarray}
H_i^x&=&J \sum_{NN} S_j^x+ D(S_{i+y}^z-S_{i-y}^z) \label{LF1} \\
H_i^y&=&J \sum_{NN} S_j^y- D(S_{i+x}^z-S_{i-x}^z) \label{LF2}\\
H_i^z&=&J \sum_{NN} S_j^z+ D(S_{i+x}^y-S_{i-x}^y)- D(S_{i+y}^x-S_{i-y}^x)\nonumber\\
&&+H \label{LF2}
\end{eqnarray}

To determine the ground state (GS), we minimize the energy of each spin, one after another. This can be numerically achieved as the following. At each spin, we calculate its local-field components acting on it from its NN using the above equations. Next we align the spin in its local field, i. e. taking $S_i^x=H_i^x/\sqrt{H_i^x**2+H_i^y**2+H_i^z**2}$ etc. The denominator is the modulus of the local field. In doing so, the spin modulus is normalized to be 1. As seen from Eq. (\ref{LF}), the energy of the spin $\vec S_i$ is minimum. We take another spin and repeat the same procedure until all spins are visited.  This achieves one iteration. We have to perform a large number of iterations until the convergence of the system energy.  The reader is referred to Ref. \cite{Sahbi2018} for details.

\begin{figure}[h!]
\center
\includegraphics[width=7cm]{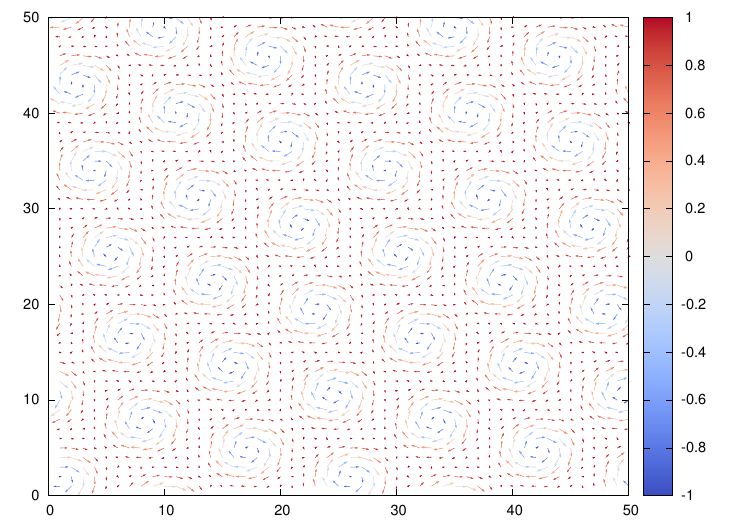}
\includegraphics[width=7cm]{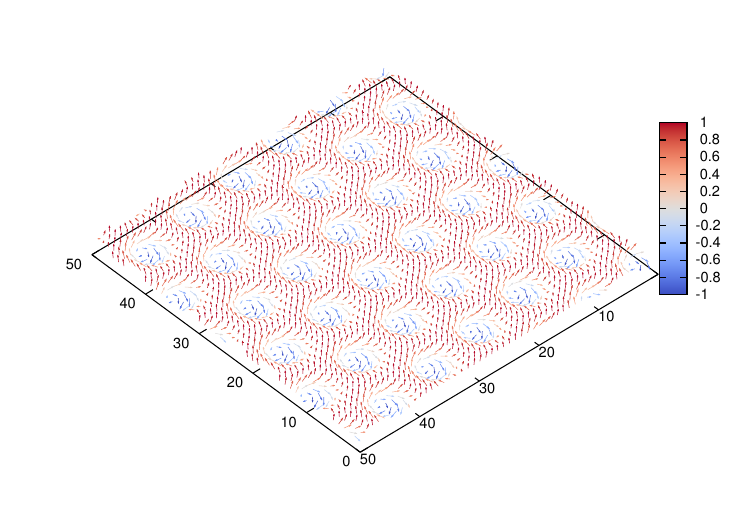}
\includegraphics[width=7cm]{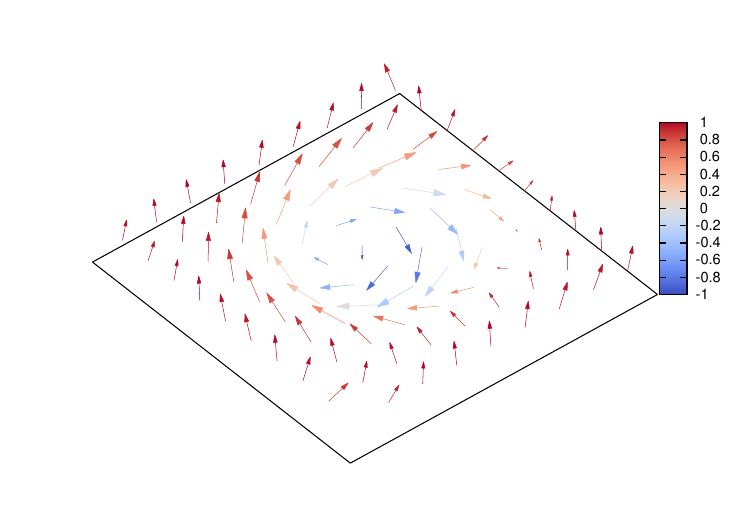}
\caption{Ground state for $D/J=1$ and $H/J=0.5$, a crystal of skyrmions is observed (a) Skyrmion crystal viewed in the $xy$ plane, (b) a 3D view, (c)  zoom of the structure of a single skyrmion. The value of $S_z$ is indicated on the color scale. See text for comments. \label{GS1}}
\end{figure}

We show in Fig. \ref{GS1} an example  calculated with  $D/J=1$ and $H/J=0.5$. Figure \ref{GS1} shows the 2D view of the GS where one sees that skyrmions form a crystal on a triangular arrangement. Fig. \ref{GS1}b shows a 3D view and Fig. \ref{GS1}c shows the spin structrure of a single  skyrmion where one sees that the spin at the center  is pointing down, and the spins at the skyrmion boundary are pointing in the +$z$ direction.   This kind of skyrmion is of Bloch-type in which the spins rotate in the tangential planes, namely perpendicular to the radial directions, when moving from the core to the periphery (the second kind is the Neel-type skyrmion in which the spins rotate in the radial planes from the core to the periphery) \cite{WangKang}. 
Note that the size of skyrmion decreases with increasing $H$.

\begin{figure}[h!]
\center
\includegraphics[width=6cm]{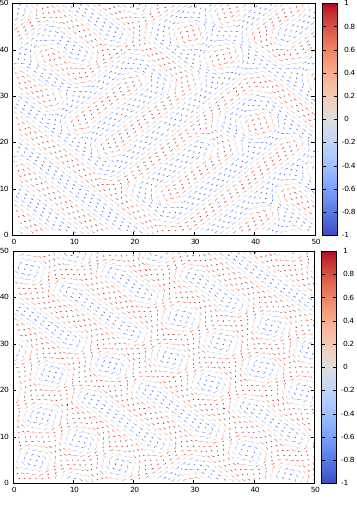}
\caption{Top: Ground state for $D/J=1$ and $H/J=0$, a mixing of domains of long and round islands, Bottom: Ground state for $D/J=1$ and $H/J=0.25$, a mixing of domains of long islands and vortices. We call these structures the "labyrinth phase". \label{GS2}}
\end{figure}

We show in Fig. \ref{GS2} the case of weak fields:  Fig. \ref{GS2}a displays a GS  at $H=0$ where domains of long and round islands of up spins separated by labyrinths of down spins.  When $H$ is increased, vortices begin to appear. The GS is a mixing of long islands of up spins and vortices as seen in Fig. \ref{GS2}b obtained with $D=1$ and $H=0.25$.  This phase can be called "labyrinth phase" or "stripe phase".

In Ref. \cite{Sahbi2018} we have showed that the  skyrmion crystal is stable at finite temperature. To see the skyrmion stability, we have defined a new order parameter as follows:

\begin{equation}\label{OP}
M(T)=\frac{1}{N^2(t_a-t_0)}\sum_i |\sum_{t=t_0}^{t_a} \mathbf S_i (T,t)\cdot \mathbf S_i^0(T=0)|
\end{equation}
where $\mathbf S_i (T,t)$ is the $i$-th spin at the time $t$, at temperature $T$, and $\mathbf S_i (T=0)$ is its state in the GS. The order parameter $M(T)$ is close to 1 at very low $T$ where each spin is only weakly deviated from its state in the GS. $M(T)$ is zero when every spin strongly fluctuates in the paramagnetic state.
The above definition of $M(T)$ is similar to the Edwards-Anderson order parameter used to measure the degree of freezing in spin glasses \cite{Mezard}: we follow each spin with time evolving and take its time averaging before taking the spatial average on all spins at the end.

\begin{figure}[htb]
\center
\includegraphics[width=8cm]{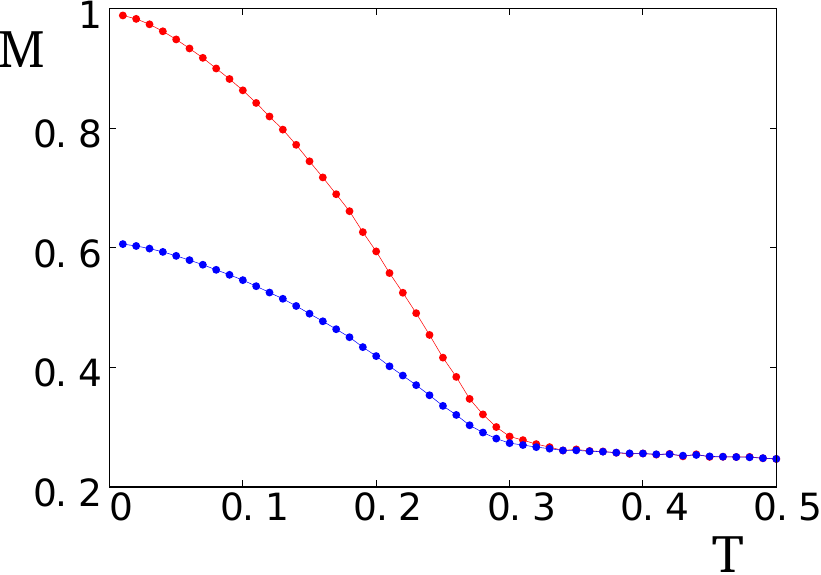}
\caption{Order parameter defined in Eq. (\ref{OP}) versus $T$ (red), for $H=0.5$ and lattice size $1800^2$, averaged during $t_a=10^5$ MC steps per spin after an equilibrating time $t_0=10^5$ MC steps. The projection
of the $S_z (T)$ on $S_z^0$ of the GS are shown in blue. See text for comments. \label{OPT}}
\end{figure}

We show in Fig. \ref{OPT} the order parameter $M$ versus $T$ (red data points) as well as the average $z$ spin component (blue data points) calculated by the projection procedure for the total time $t=10^5+10^6$ MC steps per spin.  As seen, both two curves indicate a phase transition at $T_c\simeq 0.26 J/k_B$. The fact that $M$ does not vanish above $T_c$ is due to the effect of the applied field. 

The stability of skyrmions at high temperatures is very important for making devices.  Nowadays, skyrmions have been observed at room temperatures in various materials (see for example Refs. \cite{Yu2,Gilbert}).

To conclude this section, note that we have also studied a number of issues concerning applications of skymions. To keep the length of this chapter  reasonable, we do not develop these works of us here, we just mention a few of them in Ref. \cite{XCZhang2020} to Ref.\cite{Koibuchi2024}. The reader interested in the subject of skyrmions may have a look at  these papers.

\section{Concluding remarks}\label{Concl}

This chapter gives an overview on some important properties of magnetic thin films obtained mostly from statistical physics and computer simulations by the author and his collaborators. On each item many references have been cited in addition to those cited in our original papers.  We have discussed the effect of surface spin waves on macroscopic thermodynamic properties in thin films such as the low surface magnetization and the surface phase transition. Several examples have been shown.  The criticality of the phase transition in thin films has  also been analyzed with high-performance numerical simulation methods such as the multi-histogram technique and the Wang-Landau algorithm.   The reader is referred to our cited original papers for the details of these complicated numerical methods and also the analytical Green's function theory. 

Many topics have not been reviewed above, they include spin transport  in magnetically-ordered thin films \cite{HoangDiep}, topological Hall effects \cite{Kurumaji,Gobel,Hamamoto} and numerous experimental  works on real materials.

\section*{Acknowledgments}
\addcontentsline{toc}{section}{Acknowledgements}

This chapter is a review based on subjects that the author has investigated mostly with his doctorate students over the years. The author is thankful to them, in particular R. Quartu, V.-T. Ngo, D.-T. Hoang, X.-T. Pham Phu, S. El Hog and I. Sharafullin, for fruitfull collaborations.  

%
%
%

\end{document}